\documentclass[twocolumn,showpacs,preprintnumbers,amsmath,amssymb]{revtex4}

\usepackage{graphicx}
\usepackage{dcolumn}
\usepackage{bm}

\begin{document}

\preprint{APS/123-QED}

\title{Implementation of a two-qubit controlled-U gate based on unconventional geometric phase with a constant gating time 
}

\author{B. F. C. Yabu-uti}
 \email{yabuuti@ifi.unicamp.br}
\author{J. A. Roversi}%
 \email{roversi@ifi.unicamp.br}
\affiliation{%
Instituto de F\'isica "Gleb Wataghin", Universidade Estadual de
 Campinas, 13083-859, Campinas, SP, Brazil}%

\date{\today}


\begin{abstract}

We propose an alternative scheme to implement a two-qubits Controlled-U gate in the hybrid system atom-$CCA$ (coupled cavities array). 
Our scheme results in a constant gating time and, with an adjustable qubit-bus coupling (atom-resonator), one can specify a particular transformation $U$ on the target qubit. 
We believe that this proposal may open promising perspectives for networking quantum information processors and implementing distributed and scalable quantum computation.

\end{abstract}

\pacs{Valid PACS appear here}
\maketitle

\section{\label{sec:level1}Introduction}

For a distributed quantum information processing in a quantum computer with practical applications, the coupling between different sub-systems (a large ensemble of qubits) is essential for realizing an efficient quantum communication and for implementing controllable and distributed quantum gates.

Cavity quantum electrodynamics systems (cQED), which combine atomic and photonic quantum bits, have attracted much attention because of its low decoherence rate and promising feasibility to scale up. Coupled cavities array has the advantage of easily addressing individual lattice sites with optical lasers. Furthermore, the atoms trapped in the resonators may have relatively long-lived atomic levels for encoding quantum information.

Schemes have been proposed for quantum communication \cite{pellizzari.5242} and generation of maximally entangled states \cite{yabuuti.1021} between two atoms trapped in distant optical cavities connected by an optical fiber. 

Furthermore, quantum logic gates \cite{pellizzari.3788} based on cavity QED system have been extensively investigated over the recent years. In particular, the scheme proposed by Zheng and Guo \cite{zheng.2392} has already been realized experimentally with a micromazer cavity and long-lived Rydberg atoms \cite{osnaghi.037902}.

In this work, we investigated the implementation of a two-qubits Controlled-U gate in the hybrid system atom-$CCA$ (coupled cavities array). The proposal is based on single qubit operations and an unconventional geometric phase on two identical three-level atoms, strongly driven by a resonant classical field \cite{solano.027903,chen.032344}, trapped in distant cavities connected by an optical fiber. Our scheme results in a constant gating time (which depends on the experimental parameters) and, with an adjustable qubit-bus coupling (atom-resonator), one can specify a particular transformation $U$ on the target qubit.

\section{\label{sec:level2}Basic Theory}

Before the controlled-$U$ gate implementation, it is worth discussing some of the basic theories.

\subsection{\label{sec:level2A}$CCA$ and the strongly driven Jaynes-Cummings model}

\begin{figure}[h*]
 \resizebox{\columnwidth}{!}
  {\includegraphics[width=0.35\textwidth]{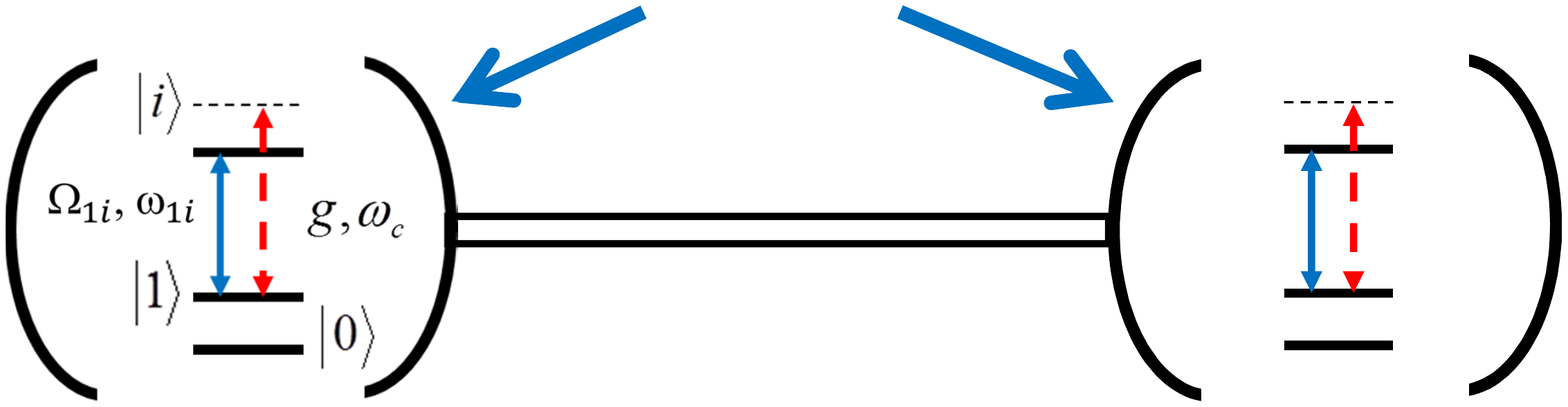}}
   \vspace{-4cm}
		\caption{Two distant atoms (and level configuration) in separate cavities connected by an optical fiber.}
 \label{fig:coupledcav}
\end{figure}

Consider two identical three-levels atoms trapped in distant cavities connected by an optical fiber, as shown in Fig.\ref{fig:coupledcav}.

In the short fiber limit the coupling between the cavity modes and fiber is given by the Hamiltonian \cite{pellizzari.5242},
\begin{equation}
	H_{cf}= \hbar\nu\hat{b}\left(\hat{a}_{1}^{\dagger}+e^{i\varphi}\hat{a}_{2}^{\dagger}\right)+ h.c,
	\label{eq:HC1FC2}
\end{equation}
\noindent where $\hat{b}$ is the annihilation operator for the fiber mode, $\hat{a}_{j}^{\dagger}$ is
the creation operator for the $j$th cavity mode, $\nu$ is the cavity fiber coupling strength, and $\varphi$ is a phase due to propagation of the field through the fiber.

Each atom has one excited (intermediate) state $\left|i\right\rangle$ and two ground states $\left|1\right\rangle$ and $\left|0\right\rangle$ (the logical qubits). The transition $\left|i\right\rangle\leftrightarrow\left|1\right\rangle$ (frequency $\omega_{1i}$) is coupled to the cavity mode with the coupling constant $g$ and detuning $\delta=\omega_{c}-\omega_{1i}$. Furthermore, the same transition is driven by a resonant classical field with Rabi frequency $\Omega_{1i}$. As the state $\left|0\right\rangle$ is not affected during the interaction, the atom-field interaction in cavity $j$ in the interaction picture is described by the Hamiltonian,
\begin{equation}
	H_{acj}= \hbar\left(g\hat{a}_{j}e^{-i\delta t} + \Omega_{1i}\right)\sigma_{+j}+ h.c,
	\label{eq:HAC}
\end{equation}
\noindent where $\sigma_{+j}=\left|i\right\rangle_{jj}\left\langle 1\right|$.

Let us consider the normal modes $\hat{c}=1/\sqrt{2}(\hat{a}_{1}-e^{i\varphi}\hat{a}_{2})$ and $\hat{c}_{\pm}=1/2(\hat{a}_{1}+e^{i\varphi}\hat{a}_{2}\pm\sqrt{2}\hat{b})$ with frequency $\omega_{c}$ and $\omega_{c}\pm\sqrt{2}\nu$. The whole Hamiltonian in the interaction picture can be rewritten as,
\begin{equation}
	H=H_{0}+H_{1}, 
	\label{eq:H}
\end{equation}
\noindent where,
\begin{equation}
	H_{0}=\hbar\sqrt{2}\nu c_{+}^{\dagger}c_{+}-\hbar\sqrt{2}\nu c_{-}^{\dagger}c_{-}+\hbar\Omega_{1i}\sum_{j=1}^{2}\left(\sigma_{+j}+\sigma_{-j}\right),
	\label{eq:H0}
\end{equation}
\begin{eqnarray}
	H_{1}= \hbar g\left[\frac{1}{2}\left(\hat{c}_{+}+\sqrt{2}\hat{c}+\hat{c}_{-}\right)\sigma_{+1}e^{-i\delta t}\right.\\\nonumber\left.+\frac{1}{2}\left(\hat{c}_{+}-\sqrt{2}\hat{c}+\hat{c}_{-}\right)\sigma_{+2}e^{-i\delta t}+h.c\right].
	\label{eq:H1}
\end{eqnarray}

We now switch to a new atomic basis $\left|+\right\rangle_{j}=1/\sqrt{2}(\left|1\right\rangle_{j}+\left|i\right\rangle_{j})$ and $\left|-\right\rangle_{j}=1/\sqrt{2}(\left|1\right\rangle_{j}-\left|i\right\rangle_{j})$ and perform the unitary transformation $U=e^{-iH_{0}t/\hbar}$, that results in \cite{solano.027903,chen.032344}, 
\begin{eqnarray}
\nonumber
 H'=U^{\dagger}H_{1}U=~~~~~~~~~~~~~~~~~~~~~~~~~~~~~~~~~~~~~~~~~~~~~~~~~~~\\\nonumber
 \hbar g\left[\frac{1}{2}\left(\hat{c}_{+}e^{-i(\delta+\sqrt{2}\nu)t}+\sqrt{2}\hat{c}e^{-i\delta t}+\hat{c}_{-}e^{-i(\delta-\sqrt{2}\nu)t}\right)\right.\\\nonumber
\left.\times\frac{1}{2}\left(\tilde{\sigma}_{z1}-\tilde{\sigma}_{-1}e^{-i2\Omega_{1i}t}+\tilde{\sigma}_{+1}e^{i2\Omega_{1i}t}\right)\right.~~~~~~~~~~\\\nonumber
\left.+\frac{1}{2}\left(\hat{c}_{+}e^{-i(\delta+\sqrt{2}\nu)t}-\sqrt{2}\hat{c}e^{-i\delta t}+\hat{c}_{-}e^{-i(\delta-\sqrt{2}\nu)t}\right)\right.\\\left.\times\frac{1}{2}\left(\tilde{\sigma}_{z2}-\tilde{\sigma}_{-2}e^{-i2\Omega_{1i}t}+\tilde{\sigma}_{+2}e^{i2\Omega_{1i}t}\right)+h.c\right]~~~~,
	\label{eq:Hx}
\end{eqnarray}
\noindent where $\tilde{\sigma}_{zj}\left|\pm\right\rangle_{j}=\pm\left|\pm\right\rangle_{j}$ and $\tilde{\sigma}_{\pm j}\left|\mp\right\rangle_{j}=\left|\pm\right\rangle_{j}$.

For a strong cavity-fiber coupling $\nu>>g$ and intense driving regime $\Omega>>g,\delta$, we can neglect the terms oscillating fast,
\begin{equation}
	H'_{eff}= \frac{\hbar g}{2\sqrt{2}}\left(\hat{c}e^{-i\delta t}+\hat{c}^{\dagger}e^{i\delta t}\right)\left(\tilde{\sigma}_{z1}-\tilde{\sigma}_{z2}\right).
	\label{eq:Heff}
\end{equation}

The evolution operator for Hamiltonian (\ref{eq:Heff}) can be written as \cite{zheng.060303},
\begin{equation}	U'=e^{-iA(t)(\tilde{\sigma}_{z1}-\tilde{\sigma}_{z2})^{2}}e^{-iB(t)\hat{c}(\tilde{\sigma}_{z1}-\tilde{\sigma}_{z2})}e^{-iB^{*}(t)\hat{c}^{\dagger}(\tilde{\sigma}_{z1}-\tilde{\sigma}_{z2})},
\end{equation}
\noindent in order to find the time dependent functions $A$ and $B$ we can use the Schrödinger equation and obtain,
\begin{eqnarray}
B(t)=-\frac{g}{2\sqrt{2}i\delta}\left(e^{-i\delta t}-1\right),
\\\nonumber
A(t)=-\frac{g^{2}}{8\delta}\left(t-\frac{1}{i\delta}\left(e^{i\delta t}-1\right)\right).
\end{eqnarray}
 
When the interaction time satisfies $t=\tau=2\pi/\delta$, the whole evolution operator of the system can be expressed as,
\begin{equation}	U(\tau)=e^{-iH_{0}\tau/\hbar}U'(\tau)=e^{-i\Omega_{1i}\tau(\tilde{\sigma}_{z1}+\tilde{\sigma}_{z2})}e^{i\lambda\tau(\tilde{\sigma}_{z1}-\tilde{\sigma}_{z2})^{2}},
\end{equation}
\noindent with $\lambda=g^{2}/8\delta$.

It is evident that such an evolution operator is independent of the cavity mode, which means that the atoms get insensitive to the thermal fluctuation.

\subsection{\label{sec:level2B}Microwave and optical pulses}

Consider a particular atomic transition $\left|i\right\rangle\leftrightarrow\left|j\right\rangle$ ($\left|i\right\rangle$ is the lower energy level) driven by a resonant classical (optical or microwave) pulse. The interaction Hamiltonian in the interaction picture is then given by,
\begin{equation}
	H=\hbar(\Omega_{ij}e^{i\phi}\left|i\right\rangle\left\langle j\right|+h.c),
	\label{eq:hpulse}
\end{equation}
\noindent in which $\Omega_{ij}$ and $\phi$ are the rabi frequency and the initial phase of the pulse, respectively. From the Hamiltonian (\ref{eq:hpulse}) it is easy to find the following state rotation due a pulse of duration $t$ \cite{scully},
\begin{eqnarray}
\left|i\right\rangle=cos\Omega_{ij}t\left|i\right\rangle-ie^{-i\phi}sin\Omega_{ij}t\left|j\right\rangle,
\\\nonumber
\left|j\right\rangle=cos\Omega_{ij}t\left|j\right\rangle-ie^{i\phi}sin\Omega_{ij}t\left|i\right\rangle.~~
\end{eqnarray}

\section{\label{sec:level3}Controlled-U Gate implementation}

Previously we had introduced two types of interactions of qubit systems with a cavity mode and/or pulses. Here such a background will be employed for the gate implementation.

Initially the atoms are in one of the computational basis states $\left\{\left|0\right\rangle_{1}\left|0\right\rangle_{2},\left|0\right\rangle_{1}\left|1\right\rangle_{2},\left|1\right\rangle_{1}\left|0\right\rangle_{2},\left|1\right\rangle_{1}\left|1\right\rangle_{2}\right\}$ 
(the first qubit is the control and the second one is the target qubit) and the $CCA$ system is prepared in the vacuum state $\left|0\right\rangle_{c}$. Indeed, in the ideal case, the $CCA$ system can be in any state. 

We propose that a controlled-$U$ gate can be implemented through the following operations:
\begin{itemize}
	\item STEP 1: Apply a microwave pulse (with a frequency $\omega_{01}$ and $\phi=-\pi/2$) in the target qubit for $\Omega_{01}t_{1}=\pi/4$. Such a single qubit operation create a superposition state,
\begin{eqnarray}
	\nonumber
	\left|0\right\rangle_{2}\rightarrow\frac{1}{\sqrt{2}}\left(\left|0\right\rangle_{2}+\left|1\right\rangle_{2}\right),
	\\
	\left|1\right\rangle_{2}\rightarrow\frac{1}{\sqrt{2}}\left(-\left|0\right\rangle_{2}+\left|1\right\rangle_{2}\right),
\end{eqnarray}
	\item STEP 2: Apply an optical pulse (with a frequency $\omega_{1i}$ and $\phi=\pi/2$) in both qubits for $\Omega_{1i}t_{2}=\pi/4$. After the pulse, we have $\left|1\right\rangle_{j}\rightarrow\left|-\right\rangle_{j}$.
	\item STEP 3: Turn on the atom-field interaction as described in section \ref{sec:level2A} for $t_{3}=\tau=2\pi/\delta$ and $\Omega_{1i}=50\delta$,
\begin{eqnarray}
	\nonumber
\left|0\right\rangle_{1}\left|0\right\rangle_{2}\left|0\right\rangle_{c}\rightarrow\left|0\right\rangle_{1}\left|0\right\rangle_{2}\left|0\right\rangle_{c},~~~~~~~~
	\\
	\nonumber
\left|0\right\rangle_{1}\left|-\right\rangle_{2}\left|0\right\rangle_{c}\rightarrow e^{i\lambda\tau}\left|0\right\rangle_{1}\left|-\right\rangle_{2}\left|0\right\rangle_{c},
	\\
	\nonumber
\left|-\right\rangle_{1}\left|0\right\rangle_{2}\left|0\right\rangle_{c}\rightarrow e^{i\lambda\tau}\left|-\right\rangle_{1}\left|0\right\rangle_{2}\left|0\right\rangle_{c},
	\\
\left|-\right\rangle_{1}\left|-\right\rangle_{2}\left|0\right\rangle_{c}\rightarrow\left|-\right\rangle_{1}\left|-\right\rangle_{2}\left|0\right\rangle_{c},~~~~~~
\end{eqnarray}	
	
	\item STEP 4: Repeat the operation of step (2) which now results in $\left|-\right\rangle_{j}\rightarrow -\left|i\right\rangle_{j}$,
	\item STEP 5: Apply an optical pulse (with a frequency $\omega_{1i}$ and $\phi=-\lambda\tau-\pi/2$) in both qubits for $\Omega_{1i}t_{4}=\pi/2$ which results,
\begin{equation}
	\left|i\right\rangle_{j}\rightarrow -e^{-i\lambda\tau}\left|1\right\rangle_{j},	
\end{equation}
	\item STEP 6: Repeat the operation of step (1) but with $\phi=\pi/2$.
\end{itemize}

The states of the two qubits (atoms) after such a procedure are,
\begin{eqnarray}
\nonumber  
\left|0\right\rangle_{1}\left|0\right\rangle_{2}\rightarrow\left|0\right\rangle_{1}\left|0\right\rangle_{2},~~~~~~~~~~~~~~~~~~~~~~
\\\nonumber  
\left|0\right\rangle_{1}\left|1\right\rangle_{2}\rightarrow\left|0\right\rangle_{1}\left|1\right\rangle_{2},~~~~~~~~~~~~~~~~~~~~~~
\\\nonumber
\left|1\right\rangle_{1}\left|0\right\rangle_{2}\rightarrow\left|1\right\rangle_{1} e^{-i\Theta(g)}(cos\Theta(g)\left|0\right\rangle_{2}-isin\Theta(g)\left|1\right\rangle_{2}),
\\
\left|1\right\rangle_{1}\left|1\right\rangle_{2}\rightarrow\left|1\right\rangle_{1} e^{-i\Theta(g)}(cos\Theta(g)\left|1\right\rangle_{2}-isin\Theta(g)\left|0\right\rangle_{2}),
\end{eqnarray}
\noindent with,
\begin{equation}
	\Theta(g)=\lambda\tau=\frac{g^{2}\pi}{4\delta^{2}},
\end{equation}
\noindent which implies that if and only if the control qubit is in the state $\left|1\right\rangle$ a unitary transformation (that can be appropriately chosen varying the atom-field coupling strength) is performed on the target qubit and nothing happens otherwise. Moreover, such an operation is implemented in a constant time given by,
\begin{equation}
	t_{tot}=t_{1}+t_{2}+t_{3}+t_{4}+t_{5}+t_{6}=\frac{\pi}{2\Omega_{01}}+\frac{\pi}{\Omega_{1i}}+\frac{2\pi}{\delta}
\end{equation}
\noindent that depends only on experimental parameters. 

\section{\label{sec:level6}Discussion}

In summary we have described a multi-step protocol for implementing a controlled-$U$ gate for two atoms in separate cavities connected by an optical fiber. 

In contrast to a previous proposal \cite{yang.032317}, our scheme is implemented in a constant gating time and, with an adjustable qubit-bus coupling (atom-resonator), one can specify a particular transformation $U$ on the target qubit.

Choosing experimental parameters such as $\Omega_{01}=10\delta$, $\Omega_{1i}=100\delta$ and $\delta\cong 1GHz$ (in agreement with previous works \cite{michael.849,peng.1207}) we can achieve a total operation time of $t_{tot}=3,3ns$ (disregarding delays between the steps), which is considerably small in comparison with to previous proposal \cite{yang.032317}.

The combination of fiber-based cavities and atom-chip technology is a promising candidatefor the implementation of our proposal. In such a system, each atom (or atom cloud) can be strongly coupled to the cavity mode and positioned deterministically anywhere within the cavity giving rise to a controlled, tunable coupling rate \cite{colombe.272} with a high single-atom cooperativity factor of $g^{2}/2\kappa\gamma=145$, where $\kappa$ is the cavity photon decay rate and $\gamma$ is the atomic spontaneous emission rate.

Two important issues deserve considerations:(i) a more rigorous study of the influence of dissipation on the proposal, especially in step $3$; (ii) an analysis of errors during the execution of the protocol, especially in step $5$. 

Even without a tunable constant coupling $g$, one can still implement a controlled-$U$ gate by varying the interaction time in step $3$ ($t_{3}=\tau_{n}=2\pi n/\delta$), but then the set of available transformation $U$ is only a discrete set.

\begin{acknowledgments}
B.F.C.Y thanks the financial support from Conselho Nacional de Desenvolvimento Cient\'ifico e Tecnol\'ogico (CNPQ). J.A.R. thanks CNPq for partial support of this work. J.A.R. also acknowledge partial support of CEPOF (Centro de Pesquisa em \'Optica e Fot\^onica).
\end{acknowledgments}


\begin{thebibliography}{13}
\expandafter\ifx\csname natexlab\endcsname\relax\def\natexlab#1{#1}\fi
\expandafter\ifx\csname bibnamefont\endcsname\relax
  \def\bibnamefont#1{#1}\fi
\expandafter\ifx\csname bibfnamefont\endcsname\relax
  \def\bibfnamefont#1{#1}\fi
\expandafter\ifx\csname citenamefont\endcsname\relax
  \def\citenamefont#1{#1}\fi
\expandafter\ifx\csname url\endcsname\relax
  \def\url#1{\texttt{#1}}\fi
\expandafter\ifx\csname urlprefix\endcsname\relax\def\urlprefix{URL }\fi
\providecommand{\bibinfo}[2]{#2}
\providecommand{\eprint}[2][]{\url{#2}}

\bibitem[{\citenamefont{Pellizzari}(1997)}]{pellizzari.5242}
\bibinfo{author}{\bibfnamefont{T.}~\bibnamefont{Pellizzari}},
  \bibinfo{journal}{Phys. Rev. Lett.} \textbf{\bibinfo{volume}{79}},
  \bibinfo{pages}{5242} (\bibinfo{year}{1997}).

\bibitem[{\citenamefont{Yabu-uti et~al.}(2008)\citenamefont{Yabu-uti, Nohama,
  and Roversi}}]{yabuuti.1021}
\bibinfo{author}{\bibfnamefont{B.~F.~C.} \bibnamefont{Yabu-uti}},
  \bibinfo{author}{\bibfnamefont{F.~K.} \bibnamefont{Nohama}},
  \bibnamefont{and} \bibinfo{author}{\bibfnamefont{J.~A.}
  \bibnamefont{Roversi}}, \bibinfo{journal}{International Journal of Quantum
  Information} \textbf{\bibinfo{volume}{6}}, \bibinfo{pages}{1021 }
  (\bibinfo{year}{2008}).

\bibitem[{\citenamefont{Pellizzari et~al.}(1995)\citenamefont{Pellizzari,
  Gardiner, Cirac, and Zoller}}]{pellizzari.3788}
\bibinfo{author}{\bibfnamefont{T.}~\bibnamefont{Pellizzari}},
  \bibinfo{author}{\bibfnamefont{S.~A.} \bibnamefont{Gardiner}},
  \bibinfo{author}{\bibfnamefont{J.~I.} \bibnamefont{Cirac}}, \bibnamefont{and}
  \bibinfo{author}{\bibfnamefont{P.}~\bibnamefont{Zoller}},
  \bibinfo{journal}{Phys. Rev. Lett.} \textbf{\bibinfo{volume}{75}},
  \bibinfo{pages}{3788} (\bibinfo{year}{1995}).

\bibitem[{\citenamefont{Zheng and Guo}(2000)}]{zheng.2392}
\bibinfo{author}{\bibfnamefont{S.-B.} \bibnamefont{Zheng}} \bibnamefont{and}
  \bibinfo{author}{\bibfnamefont{G.-C.} \bibnamefont{Guo}},
  \bibinfo{journal}{Phys. Rev. Lett.} \textbf{\bibinfo{volume}{85}},
  \bibinfo{pages}{2392} (\bibinfo{year}{2000}).

\bibitem[{\citenamefont{Osnaghi et~al.}(2001)\citenamefont{Osnaghi, Bertet,
  Auffeves, Maioli, Brune, Raimond, and Haroche}}]{osnaghi.037902}
\bibinfo{author}{\bibfnamefont{S.}~\bibnamefont{Osnaghi}},
  \bibinfo{author}{\bibfnamefont{P.}~\bibnamefont{Bertet}},
  \bibinfo{author}{\bibfnamefont{A.}~\bibnamefont{Auffeves}},
  \bibinfo{author}{\bibfnamefont{P.}~\bibnamefont{Maioli}},
  \bibinfo{author}{\bibfnamefont{M.}~\bibnamefont{Brune}},
  \bibinfo{author}{\bibfnamefont{J.~M.} \bibnamefont{Raimond}},
  \bibnamefont{and} \bibinfo{author}{\bibfnamefont{S.}~\bibnamefont{Haroche}},
  \bibinfo{journal}{Phys. Rev. Lett.} \textbf{\bibinfo{volume}{87}},
  \bibinfo{pages}{037902} (\bibinfo{year}{2001}).

\bibitem[{\citenamefont{Solano et~al.}(2003)\citenamefont{Solano, Agarwal, and
  Walther}}]{solano.027903}
\bibinfo{author}{\bibfnamefont{E.}~\bibnamefont{Solano}},
  \bibinfo{author}{\bibfnamefont{G.~S.} \bibnamefont{Agarwal}},
  \bibnamefont{and} \bibinfo{author}{\bibfnamefont{H.}~\bibnamefont{Walther}},
  \bibinfo{journal}{Phys. Rev. Lett.} \textbf{\bibinfo{volume}{90}},
  \bibinfo{pages}{027903} (\bibinfo{year}{2003}).

\bibitem[{\citenamefont{Chen et~al.}(2006)\citenamefont{Chen, Feng, Zhang, and
  Gao}}]{chen.032344}
\bibinfo{author}{\bibfnamefont{C.-Y.} \bibnamefont{Chen}},
  \bibinfo{author}{\bibfnamefont{M.}~\bibnamefont{Feng}},
  \bibinfo{author}{\bibfnamefont{X.-L.} \bibnamefont{Zhang}}, \bibnamefont{and}
  \bibinfo{author}{\bibfnamefont{K.-L.} \bibnamefont{Gao}},
  \bibinfo{journal}{Phys. Rev. A} \textbf{\bibinfo{volume}{73}},
  \bibinfo{pages}{032344} (\bibinfo{year}{2006}).

\bibitem[{\citenamefont{Zheng}(2002)}]{zheng.060303}
\bibinfo{author}{\bibfnamefont{S.-B.} \bibnamefont{Zheng}},
  \bibinfo{journal}{Phys. Rev. A} \textbf{\bibinfo{volume}{66}},
  \bibinfo{pages}{060303} (\bibinfo{year}{2002}).

\bibitem[{\citenamefont{Scully and Zubairy}(1997)}]{scully}
\bibinfo{author}{\bibfnamefont{M.~O.} \bibnamefont{Scully}} \bibnamefont{and}
  \bibinfo{author}{\bibfnamefont{M.~S.} \bibnamefont{Zubairy}},
  \emph{\bibinfo{title}{Quantum optics}} (\bibinfo{publisher}{Cambrigde Univ.
  Press}, \bibinfo{year}{1997}), \bibinfo{edition}{$1^{o}$} ed.

\bibitem[{\citenamefont{Yang and Han}(2006)}]{yang.032317}
\bibinfo{author}{\bibfnamefont{C.-P.} \bibnamefont{Yang}} \bibnamefont{and}
  \bibinfo{author}{\bibfnamefont{S.}~\bibnamefont{Han}},
  \bibinfo{journal}{Phys. Rev. A} \textbf{\bibinfo{volume}{73}},
  \bibinfo{pages}{032317} (\bibinfo{year}{2006}).

\bibitem[{\citenamefont{Hartmann et~al.}(2006)\citenamefont{Hartmann, Brandao,
  and Plenio}}]{michael.849}
\bibinfo{author}{\bibfnamefont{M.~J.} \bibnamefont{Hartmann}},
  \bibinfo{author}{\bibfnamefont{F.~G. S.~L.} \bibnamefont{Brandao}},
  \bibnamefont{and} \bibinfo{author}{\bibfnamefont{M.~B.}
  \bibnamefont{Plenio}}, \bibinfo{journal}{Nature Phys.}
  \textbf{\bibinfo{volume}{2}}, \bibinfo{pages}{849} (\bibinfo{year}{2006}).

\bibitem[{\citenamefont{Li and Li}(2011)}]{peng.1207}
\bibinfo{author}{\bibfnamefont{P.-B.} \bibnamefont{Li}} \bibnamefont{and}
  \bibinfo{author}{\bibfnamefont{F.-L.} \bibnamefont{Li}},
  \bibinfo{journal}{Opt. Express} \textbf{\bibinfo{volume}{19}},
  \bibinfo{pages}{1207} (\bibinfo{year}{2011}).

\bibitem[{\citenamefont{Colombe et~al.}(2007)\citenamefont{Colombe, Steinmetz,
  Dubois, Linke, Hunger, and Reichel}}]{colombe.272}
\bibinfo{author}{\bibfnamefont{Y.}~\bibnamefont{Colombe}},
  \bibinfo{author}{\bibfnamefont{T.}~\bibnamefont{Steinmetz}},
  \bibinfo{author}{\bibfnamefont{G.}~\bibnamefont{Dubois}},
  \bibinfo{author}{\bibfnamefont{F.}~\bibnamefont{Linke}},
  \bibinfo{author}{\bibfnamefont{D.}~\bibnamefont{Hunger}}, \bibnamefont{and}
  \bibinfo{author}{\bibfnamefont{J.}~\bibnamefont{Reichel}},
  \bibinfo{journal}{Nature} \textbf{\bibinfo{volume}{450}},
  \bibinfo{pages}{272} (\bibinfo{year}{2007}).

\end{thebibliography}
\end{document}